\documentclass{aastex63}
\usepackage[caption=false]{subfig}
\usepackage{multirow}
\usepackage{graphicx}
\usepackage{booktabs}
\usepackage{amsmath,amssymb}
\usepackage[T1]{fontenc}

\graphicspath{{./}{figures/}}

\begin{document}
\title{GRB 101225A as Orphan Dipole Radiation of a Newborn Magnetar with Precession Rotation in an Off-Axis Gamma-Ray Burst}
\correspondingauthor{En-Wei Liang}
\email{lew@gxu.edu.cn}
\author{Le Zou}
\affil{Guangxi Key Laboratory for Relativistic Astrophysics, Department of Physics, Guangxi University, Nanning 530004, China; lew@gxu.edu.cn}
\author{Tian-Ci Zheng}
\affil{Guangxi Key Laboratory for Relativistic Astrophysics, Department of Physics, Guangxi University, Nanning 530004, China; lew@gxu.edu.cn}
\author{Xing Yang}
\affil{Guangxi Key Laboratory for Relativistic Astrophysics, Department of Physics, Guangxi University, Nanning 530004, China; lew@gxu.edu.cn}
\author{Hai-ming Zhang}
\affil{School of Astronomy and Space Science, Nanjing University, Nanjing 210023, China}
\author{Xiao-Yan Li}
\affil{Guangxi Key Laboratory for Relativistic Astrophysics, Department of Physics, Guangxi University, Nanning 530004, China; lew@gxu.edu.cn}
\author{Jia Ren}
\affil{School of Astronomy and Space Science, Nanjing University, Nanjing 210023, China}
\affil{Guangxi Key Laboratory for Relativistic Astrophysics, Department of Physics, Guangxi University, Nanning 530004, China; lew@gxu.edu.cn}
\author{Da-Bin Lin}
\affil{Guangxi Key Laboratory for Relativistic Astrophysics, Department of Physics, Guangxi University, Nanning 530004, China; lew@gxu.edu.cn}
\author{En-Wei Liang*}
\affil{Guangxi Key Laboratory for Relativistic Astrophysics, Department of Physics, Guangxi University, Nanning 530004, China; lew@gxu.edu.cn}

\begin{abstract}
The unusual multiwavelength lightcurves of GRB 101225A are revisited by assuming that it is from an off-axis GRB powered by a newborn magnetar. We show that its optical afterglow lightcurve is fitted with the forward shock model by parameterizing its jet structure as a Gaussian function with a half opening angle of the jet core as $1.67^{\rm o}$. The derived initial Lorentz factor ($\Gamma_0$) is 120, and the viewing angle to the jet axis is $\theta_v=3.7^{\rm o}$. Tentative QPO signatures of $P=488$ seconds and $P=250\sim 300$ seconds are found with a confidence level of 90\% by analysing its X-ray flares observed in the time interval of $[4900,\ 7500]$ seconds. Its global gamma-ray/X-ray lightcurve and the QPO signatures are represented with the magnetar dipole radiation (DR) model by considering the magnetar precession motion, assuming that the magnetar spindown is dominated by the GW emission. The bulk Lorentz factor of the DR ejecta is limited to 8, being much lower than $\Gamma_0$. Comparing GRB 101225A with the extremely off-axis GRB 170817A, we suspect that the nature of the two-component jet in GRB 170817A is a combination of a co-axial GRB jet and a DR ejecta. GRB 101225A would be among the brightest ones of the CDF-S XT2 like X-ray transient population driven by newborn magnetars. Discussion on detectability of its gravitational wave emission is also presented.
\end{abstract}


\section{Introduction} 

A rapidly spinning, strongly magnetized neutron star, the so-called ``millisecond magnetar'', could be served as the energy source of gamma-ray bursts (GRBs) (Usov 1992; Dai \& Lu 1998; Zhang \& M\'{e}sz\'{a}ros 2001; Metzger et al. 2011). The discovery of association among GRB, gravitational wave (GW) event, and kilonova of GRB170817A/GW 170817/AT2017gfo from the merger of neutron stars not only opens the multi-messenger era of GRB study but also renews our understanding of the GRB physics (Abbott et al. 2017a, 2017b). It was shown that GRB 170817A should be from off-axis observations to a structured jet (Margutti et al. 2017; Takahashi \& Ioka 2021) and its central engine might be a magnetar (Metzger et al. 2018). A newborn magnetar loses its rotational energy via the electromagnetic (EM) dipole radiation (DR) and/or GW emission. Its EM lightcurve is featured as a steady, long-lasting emission segment with a transition to a decay phase as $t^{-\beta}$ at a characteristic timescale ($\tau_{\rm SD}$), where $\beta={-1}$ if GW emission dominates the spindown losses and $\beta={-2}$ if EM losses are dominant instead (e.g. Dai \& Lu 1998; Zhang \& M\'{e}sz\'{a}ros 2001; L\"{u} et al. 2018). A rapid drop of the X-ray flux following the plateau is observed in some cases (Troja et al. 2007; Liang et al. 2007; L\"{u} et al. 2014), which is regarded as evidence of a supra massive magnetar collapsing to a black hole (Troja et al. 2007). In such a jet-wind configuration, its prompt gamma-ray emission may not be able to trigger instruments or could be observed as a low-luminosity GRB, but orphan DRs or jet afterglows may be detectable for an off-axis observer. Multiple messenger observations of GRB 170817A favor such a paradigm. Xue et al. (2019) discovered such a candidate from the deep X-ray survey by the Chandra X-ray telescope. By assuming a quasi-universal Gaussian jet, Xie et al. (2020) illustrated that the observed GRBs from their jets and the X-ray emission from their DR winds powered by newborn magnetars can be reproduced via the Monte Carlo simulations.

The DR would be a probe for the evolution of a newborn magnetar. At birth of a magnetar, the spin axis $\boldsymbol{\Omega}$ is misaligned to the symmetry axis of the magnetic field $\textbf{B}$ with a tiny inclination angle $\alpha$. The internal viscous dissipation will drive the magnetic symmetry axis orthogonal to the spin axis, which could cause a fast growth of $\alpha$, causing a strong GW signal (Cutler 2002; Stella et al. 2005; Dall'Osso \& Stella 2007). In the early stages, the effect of the internal viscous dissipation would largely overcome the effect of the electromagnetic (EM) torques. As the magnetar spins down, the EM torques take over and dominate the $\alpha$ evolution and the magnetar precession (Goldreich 1970; Jones 1976; Dall'Osso et al. 2009; Philippov et al. 2014; Dall'Osso et al. 2015; \c{S}a\c{s}maz Mu\c{s} et al. 2019). This is caused by its rapidly rotational and highly elliptical structure (Goldreich 1970; Spitkovsky 2006; Philippov et al. 2014; Zanazzi \& Lai 2015; \c{S}a\c{s}maz Mu\c{s} et al. 2019; Suvorov \& Kokkotas 2020). In the case of GW emission dominates the spin down, $\alpha$ cannot be equal to 0. Observationally, the precession may lead to the oscillations of the observed X-ray flux from its DR (e.g. Suvorov \& Kokkotas 2020).

GRB 101225A is a long GRB at redshift $z = 0.847$ (Levan et al. 2014). It is interesting due to its extremely unusual gamma-ray and X-ray lightcurves. Its gamma-ray lightcurve shows a plateau up to 1700 s post the $Swift$ Burst Alert Telescope (BAT) trigger (Cummings \& Sakamoto 2010). The gamma-ray plateau is followed by X-ray flares whose peak fluxes decay as $F_X\propto t^{-1}$ (L\"{u} et al. 2018). The X-ray emission rapid drops with a decay slope of $-6$ at $t>2.5\times 10^{4}$ s. This GRB was claimed as an `ultra-long' GRB of a unique population from different progenitors (Levan et al. 2014), such as a tidal disruption event of a minor body around an isolated Galactic neutron star (Campana et al. 2011), a merger of a helium star with a neutron star (Th\"{o}ne et al. 2011). L\"{u} et al. (2018) proposed that GRB 101225A is driven by the DR of a newborn magnetar when its rotation energy is lost via the GW emission before its collapses to a black hole (see also Zou et al. 2019). The temporal feature of its optical afterglows is completely different from the X-ray emission. It presents a plateau up to $\sim 2\times 10^{5}$ s, then decays as $t^{-1.3}$ (Campana et al. 2011).

We revisit the nature of GRB 101225A motivated by the discovery of off-axis GRB 170817A and CDF-S XT2. We propose that this GRB is a typical GRB powered by a newborn magnetar viewing off-axis to the GRB jet, being similar to GRB 170817A. In Section 2, we model the orphan optical afterglow lightcurve and estimate the viewing angle with respect to the jet axis in the structured jet scenario. In Section 3, we investigate X-ray flares as signature of the magnetar precession in GRB 101225A. Our discussions and conclusions are presented in Section 4. Throughout, a concordance cosmology with parameters $H_{0} = 70 \text{ km s}^{-1} \text{ Mpc}^{-1}$, $\Omega_{M}=0.3$, and $\Omega_{\Lambda}=0.7$ is adopted.

\section{Orphan Optical Afterglows of Off-Axis GRB 101225A}

Fig.\ref{1} shows the multiple lightcurves of GRB 101225A. One can observe that the X-ray and optical lighcurves are completely different. The highly variable X-ray flares with a sharp drop at $2.5\times 10^4$ s, which indicates that the X-ray emission is from an internal energy dissipation process. The optical flux keeps almost constant, then decay as $t^{-1.3}$ at $t>2\times 10^{5}$ s. We propose that the optical emission results from off-axis observation of a structure jet. We model the optical data with the standard external shock model by parameterizing the jet structure as a Gaussian jet (Ren et al. 2020), i.e.

\begin{equation}\label{Eq:E_k}
{E_{{\rm{k}},{\rm{iso}}}}(\theta)= {E_{{\rm{k}},{\rm{iso}},{\rm{c}}}}\exp [ - \frac{1}{2}{(\frac{\theta} {\theta _{\rm c}})^{\rm{2}}}]\  \  \   \   \ (\theta<\theta_j),
\end{equation}
where $E_{\rm k, iso, c}$ is the kinetic isotropic energy ($E_{\rm k, iso}$) observed within the characteristic half opening angle $\theta_c$. The medium profile surrounding the jet is taken as an inter-stellar medium (ISM). The physical origin of the last X-ray data, $F_{\rm X}=2.34\times 10^{-13}$ erg cm$^{-2}$ s$^{-1}$ at $t=5.6\times 10^{4}$ s, is uncertain. We take it as an upper limit of the X-ray afterglow. We employ the Markov Chain Monte Carlo technique to evaluate the likelihoods of the model parameters. Our best fit to the data is shown in Fig.\ref{1}. It is found that the optical data can be well represented with the following parameters, $\log E_{\rm k, iso, c}/{\rm erg}=53.81$, the initial fireball bulk Lorentz factor $\log \Gamma_{\rm 0, j}=2.03$, the electron energy partition $\log \epsilon_e=-1.10$, the magnetic energy partition $\log \epsilon_B=-3.89$, the medium density $\log n/{\rm cm^{3}}=0.12$, $p=2.42$, $\theta _{\rm c}=1.67^{\circ}$, $\theta _{\rm j}=3.00^{\circ}$, and $\theta_{v}=3.70^{\circ}$. Note that $\theta _{\rm c}$ represents the half-opening angle of energetic jet core, and $\theta _{\rm j}$ is the outer boundary of the Gaussian structured jet. It is found that $E_{\rm k, iso, c}$ of the GRB is comparable to the bright GRBs (e.g. Zou et al. 2019), and the viewing angle is slightly larger than the jet opening angle.

\section{X-ray flares as Signature of the Magnetar Precession in GRB 101225A}

\subsection{Quasi-Period Oscillations of the X-Ray Flares}

The X-ray lightcurve of GRB 101225A is composed of flares that have flux variation of about one order of magnitude. Tentative quasi period oscillation (QPO) feature is visually observed. We search for such a QPO feature with the well-sampled X-ray lightcurve of GRB 101225A in the time interval $[4900,\ 7500]$~s. We generated the X-ray lightcurve with a confidence level of $7\sigma$ for each data point. We adopted the Lomb-Scargle Periodogram (LSP) algorithm (Lomb 1976; Scargle 1982) to calculate the power-density spectrum (PDS) of the lightcurve. The Monte Carlo simulation technique is employed to evaluate the uncertainty of the PDS. To do so, we simulate $10^4$ mock lightcurve by considering the uncertainty of each data point is a Gaussian distribution. The uncertainty of the PDS at a given period $P$ is taken as its $1\sigma$ of the PDS distribution derived from the mock lighcurves. Fig.\ref{2}~(b) shows the PDS as a function of $P$.

We evaluate the confidence level of possible QPO signatures with Bayesian statistics assuming a background red noise (Vaughan 2005, 2010). It is found that the observed PDS peaks at $P=120,\ 250,\ 304,\ 488, \ 820,\ 1247$ seconds exceed the red noise at a confidence level of $90\%$. Screening the lightcurve with a Hanmming window, the pulses at about 250, 500, and 820 seconds are still shown in the PDS curve, albeit at a confidence level of only 68\%, while the peak at P=1247 seconds disappeared, indicating this peak is resulted from the spectral leakage due to the limit of the time series ($\sim 2500$ seconds in this analysis). We smooth the observed lightcurve with a wavelet smoothing algorithm and derive the PDS curve again, finding that the PDS peak at 120 seconds was due to the data sampling effect. Finally, because the temporal coverage of the selected light curve for our analysis is only 2500 seconds, it is not possible to claim a QPO of 820 seconds. The remaining peaks at 250, 304 and 488 s survived all of our checks and selection criteria and are found to be significant at the 68\% confidence level.

\subsection{QPO as a Signature of Magnetar Precession}

Theoretically, the precession of a magnetar should result in a QPO feature in its DR during it is spun down. The total rotation energy of a magnetar is $E_{\rm rot} = \frac{1}{2} I \Omega^{2}_{s}$, where $I$ is the moment of inertia, $\Omega_{s}= 2 \pi/ P_{\rm s}$ is the angular frequency, and $P_{\rm s}$ is the  spin period of the neutron star. The magnetar loses its rotational energy via EM emission in the co-rotating plasma scenario and GW radiation as (Zhang \& M{\'e}sz{\'a}ros 2001),

\begin{eqnarray}
-\dot{E}_{\text{rot}} = -I\Omega \dot{\Omega} &=& L_{\rm EM} + L_{\rm GW} \nonumber \\
&=& \frac{\mu^2 \Omega^4}{c^3} \lambda(\alpha)+\frac{32GI^{2}\epsilon^{2}\Omega^{6}}{5c^{5}}
\nonumber \\
&=& \frac{B^2_{p} R^6 \Omega^4}{4 c^3} \lambda(\alpha)
+\frac{32GI^{2}\epsilon^{2}\Omega^{6}}{5c^{5}},
\label{Spindown}
\end{eqnarray}
where $\mu$ is the magnetic moment, $\epsilon$ is the ellipticity, $R$ is the radius, $\dot{\Omega}$ is the time derivative of angular frequency, $B_p$ is the surface magnetic field of the magnetar at the poles, and $\lambda$ is a magnetospheric factor of the neutron star. $\lambda$ depends on the orientation of the NS as $\lambda(\alpha) = \sin^2(\alpha)$ in a pure vacuum (Goldreich \& Julian 1969). Numerical simulations of charge-filled magnetospheres suggest that $\lambda(\alpha)\simeq1 + \sin^{2}\alpha$ (Spitkovsky 2006; Kalapotharakos \& Contopoulos 2009; Philippov et al. 2015). A hybrid model from Suvorov \& Kokkotas (2020) gives $\lambda(\alpha)\simeq1+\delta \sin^{2}\alpha$, where the parameter $\delta$ quantifies the magnetospheric physics ($|\delta|\leq 1$) (Philippov et al. 2014; Arzamasskiy et al. 2015).

If the magnetar initially is non-spherical and highly elliptical, its magnetospheric torques lead to the changes of the inclination angle $\alpha$. This effect can lead to the precession of the magnetar rotation and result in the observed oscillations of the observed X-ray flux (e.g. \c{S}a\c{s}maz Mu\c{s} et al. 2019; Suvorov \& Kokkotas 2020). The $\alpha$ evolution depends on the precession of the magnetar (i.e. Goldreich 1970; Zanazzi \& Lai 2015; Suvorov \& Kokkotas 2020),

\begin{equation} \label{eq:alphadot}
\dot{\alpha} \approx k \Omega_{p}  \csc \alpha \sin(\Omega_{p} t ),
\end{equation}
where $k$ is an order-unity factor related to the other Euler angles,  and $\Omega_{p}$ is the precession velocity, which is given by $\Omega_{p}=\epsilon \Omega_{s} \cos \alpha$ assuming that the system is symmetrical. $\epsilon$ is related to the mass quadrupole moment of a newborn magnetar (Jaranowski et al. 1998). A system with a large inclination angle may have a much larger value of $\epsilon$  and/or $\Omega_{s}$ at a given $\Omega_p$ (Dall'Osso et al. 2009, 2015, 2018).

We obtain

\begin{equation} \label{eq:alpha}
\begin{aligned}
\lambda(\alpha)=1+\delta \sin^{2}\alpha & \approx 1 + \delta[1-(\cos\alpha_{0} + (k \cos(\Omega_{p} t)-k))^{2}],
\end{aligned}
\end{equation}
where $\alpha_{0}$ is the initial inclination angle. As a supra massive magnetar spins down, it would collapse into a black hole when the centrifugal force cannot hold its gravity. The magnetic ``hair'' of the magnetar has to be ejected according to the no-hair theorem of the black holes (Zhang 2014; Falcke \& Rezzolla 2014). The magnetosphere fills with plasma, and it should evolve with time. In order to keep $|\delta|\lesssim 1$, we parameterize the evolution of $\delta$ as $\delta\sim-(0.1+t/t_{c})^{\omega}$ during the life time of the magnetar, where $t_{c}$ is the collapse timescale and $\omega$ is a free parameter.

The luminosity of the DR wind in the observed frame by considering the precession, the radiation efficiency, and the Doppler boosting effects is given by

\begin{equation} \label{eq:EM}
L(\phi_v, t)= \eta L_{\rm EM}(t)\lambda(\alpha) [D(\phi_v)/D(\phi_v=0)]^3,
\end{equation}
where $\eta$ is the radiation efficiency, $\phi_v$ is the viewing angle to the DR wind, $D (\phi_v)= \gamma^{-1}[1-(1-\gamma^{-2})^{1/2}\cos\phi_v]^{-1}$ is the Doppler boosting factor, and $\gamma$ is the bulk Lorentz factor of the DR wind. Considering the eject opening angle, the isotropic luminosity in the observed frame is $L_{\rm iso}(\phi_v, t)=L(\phi_v, t)(1-\cos\theta_j)^{-1}$.

\subsection{Numerical Results}
As shown in Fig.\ref{1}, the observed gamma-ray/X-ray lightcurve of GRB 101225A is composed of a plateau ($F\propto t^{0.06}$), a normal power-law decay segment ($F\propto t^{-1.11}$), and an extremely sharp flux drop ($F\propto t^{-6.10}$). In the GRB external shock model, the decay slope of the emission flux is $\sim -1$ and transits to $\sim -2$ after the jet break. The jet break should be achromatic. As shown in Figure 1, no such achromatic break is observed in the optical band. Although the X-ray decay slope of GRB 101225A in the time interval [$2\times 10^3$, $2.5\times 10^4$]~s is consistent with the expectation of the external shock model, the decay slope post $2.5\times 10^4$ s is much steeper than $-2$, and even steeper than $-3$. This fact, together with the optical afterglow data and our analysis in \S 2, convincingly evidences that the X-rays should be from an internal energy dispassion process, but not from the external shocks of the GRB jet. The most favorable scenario is that the X-rays detected before $2.5\times 10^4$ s are from the DR of a magnetar. The sharp flux drop indicates the cease of this emission component, which would be due to the collapse of the magnetar to a black hole (L\"{u} et al. 2018).

If the spin-down of the magnetar is dominated by the GW emission, the decay slope of its DR post the characteristic timescale is $-1$, and it is $-2$ for the EM dominated scenario. The decay slope post the X-ray plateau of GRB 101225A is -1.1, being consistent with the GW-dominated scenario. In this scenario, the spin velocity and DR kinetic luminosity evolves as

\begin{equation}
\Omega_{\rm s}(t) = \Omega_{\rm s,0}(1+\frac{t}{\tau_{\rm sd}})^{-1/4}, \\
\label{Luminosity_GWEM}
\end{equation}
\begin{equation}
L_{\rm EM}(t) = L_{\rm EM,0}(1+\frac{t}{\tau_{\rm sd}})^{-1},
\label{Luminosity_GWEM}
\end{equation}
where $\Omega_{\rm s,0}$ and $L_{\rm EM,0}$ are the initial angular frequency and kinetic luminosity at $t_0$, $\tau_{\rm sd}$ is a characteristic spin-down time scale given by $\tau_{\rm sd}\sim 9.1\times10^{3}\rm s $$\times I^{-1}_{45}P^{4}_{0,-3}\epsilon^{-2}_{-3}$, in which the convention $Q = 10^x Q_x$ is adopted in cgs units.

Based on Eqs. (4)-(7), we model the gamma-ray/X-ray lightcurve prior to $2.5\times 10^4$ s. The observed isotropic gamma/X-ray luminosity ($L_{\rm }$) is calculated with $L_{\rm iso}=4\pi D^{2}_{L}F(t)(1+z)^{\Gamma-2}$, where the X-ray spectral index $\Gamma$ is 1.91. Following Xie et al. (2020), we assume that the jet and DR wind powered by the magnetar is co-axial and the maximum viewing angle of the wind is the same as the viewing angle of the afterglow jet ($3.7^{\rm o}$). The evolution of the viewing angle periodically varies from 0$^{\rm o}$ to 3.7$^{\rm o}$, which is described as $\phi_v(t)=3.7^{\rm o}\times |\sin (\Omega_p t)|$. Note that $L_{\text{EM}}$, $\eta$, and $\gamma$ are degenerate. Since the energy reservoir of the DR kinetic energy is the rotational energy of the magnetar, the saturated bulk Lorentz factor of the wind depends on the radiation efficiency and magnetization of the DR wind as shown in Xiao \& Dai (2019). They showed that saturated bulk Lorentz factor would be $300\sim 100$ if $\eta=0.01\sim 0.1$. We here adopt a moderate efficiency of 0.05. $L_{\rm EM}$ is determined by $B_{\rm p}$, $P_{\rm s,0}$, and $R$. We take $R = 1.2\times 10^{6}$ cm. The parameters of $B_{\rm p}$ and $P_{\rm s,0}$ are determined by the observed plateau luminosity and the characteristic spin-down time scale $\tau_{\rm sd}$. The ellipticity $\epsilon$, $k$, $\alpha_0$, and $\gamma$ are mainly constrained by the features of the flares. We set these as free parameters, and adjust them to represent the observed gamma-ray/X-ray lightcurve at $t<2.5\times 10^4$ s and the main feature of the PDS curve in the time interval of [4700, 7500]~s. We find that the following parameter set can visually reproduce the global lightcurve and the main PDS features of GRB 101225A, i.e. $B_{\rm p}= 7.94\times10^{13}$ Gauss, $P_{\rm s,0}= 0.98$ ms, $\omega=0.4$, $\alpha_{0} = 0.50$, $k = 0.30$, $\epsilon = 6.3\times 10^{-6}$, and $\gamma=8$.

Our numerical results are shown in Figures \ref{2} and \ref{3}. To reconcile the precession theory with the data, the theoretical lightcurve is sampled the same as the data and the uncertainties of the theoretical points are also taken as the uncertainty of the corresponding data points. We make a QPO analysis for the sampled model data. The PDS curve is shown in Fig.\ref{2}~(b). One can observe that our model can roughly reproduce the main features of the lightcurve and QPOs. The model PDS curve shows a high significant peak at $P_p=460$ seconds. It is roughly consistent with the observed QPO at $P=488$ seconds. Another theoretical PDS peak with a high significance level is at $P_p=228$ seconds, which is close to the observed weak QPO feature at $P=250$ seconds. Our theoretical model does not predict a QPO at $P=304$ seconds. Note that the observed PDS peak at $304$ seconds is quite close to the peak at 250 seconds. It is possible that the two peaks are due to the spectral spilt of the power around the QPO around $250-300$ seconds.

In our analysis, the significant flares post $10^{3}$ seconds are attributed to the precession of the magnetar. The precession period depends on $\epsilon$ as $\Omega_{p}=\epsilon \Omega_s \cos \alpha$. As the magnetar is spun down, the precession period would be longer at a later epoch, depending on the evolution of $\epsilon$, $\Omega_s$, and $\alpha$. With the parameters derived in our analysis, we obtain the initial precession period as $P_{p, 0}=177$ seconds. The $P_{\rm p,0}$ is smaller than the QPO signatures of $\sim 488$ seconds or $250\sim 300$ seconds derived from the data in the time interval of [4700, 7500] seconds, being self-consistent with the spin-down situation.
However, the increase of $P_p$ in this time interval is only $4.4\% \sim 6.7\%$ for a magnetar with a characteristic spin-down timescale of $t_{\rm sd}=2.5\times 10^4$ seconds based on Eq.(6), and the increase is even only $19\%$ within $t_{\rm sd}$ in this scenario. Thus, the rapid increase of $P_p$ from initial 177 seconds to $250\sim 300$ seconds even $\sim 488$ seconds cannot be explained if $\epsilon$ and $\alpha$ do not evolve significantly, especially the QPO of 488 seconds. If ellipticity $\epsilon$ is dominated by the magnetic stresses, it does not have a substantial decrease for making a large increase of $\Omega_p$ in such a short timescale. The rotational deformation is usually axisymmetric, and the magnetic deformations can make a quadrupolar deformation. This should lead to gravitational wave emission if the magnetic axis is inclined with respect to the spin axis (e.g. Haskell et al. 2008). Thus, one possibility is that the QPO frequency is governed by a combination of a major change of $\alpha$ plus a minor change in $\Omega_s$. Because the QPOs are observed only at $t > 10^3$~s, it could be that the inclination angle $\alpha$ grew to a large value in the earlier phase ($< 10^3$~s). This is consistent with the scenario described by Dall'Osso et al. (2009, 2018; see also Dall'Osso \& Stella 2021). For example, an increase of $\alpha$ from $0^{\rm o}$ to $45^{\rm o}$ results in a $\sim 40\%$ increase of $P_p$, which can explain the increase of $P_p$ from initial 177 seconds to $\sim 250-300$ seconds. However, multiple QPO frequencies cannot be explained in this framework. We suspect that the QPO of $P=250-300$ seconds might be a harmonic of $P=488$ seconds.

\section{Conclusions and Discussion}
We have modeled the optical afterglows of GRB 101225A with the external-forward shock model by parameterizing the jet energy structure as a Gaussian function of $E_{\rm k, iso} (\theta)/ 10^{52}{\rm erg}= 64\exp [-(\theta/ 1.67^{\rm o}) ^{2}/2]$ with an initial Lorentz factor of 120 and a half opening angle of $ 3^{\rm o}$. The derived viewing angle is $\theta_v=3.7^{\rm o}$. The geometrically-corrected jet energy is $\sim 3 \times 10^{50}$ ergs, being comparable to that of GRB 170817A. This result indicates that GRB 101225A should be an off-axis event. The global gamma-ray/X-ray lightcurve is featured as an internal plateau with transition to a decaying phase as $t^{-1.1}$ at $t=2\times 10^{3}$ seconds, being consistent with the DR of a newborn magnetar by losing its spin energy via GW emission. Analysis to the X-ray flares observed in the time interval of $[4900,\ 7500]$~s post the BAT trigger yields QPO signatures of $P=488$ seconds and $P=250\sim 300$ seconds with a confidence level of 90\%. We find that the lightcurve of GRB 101225A and its QPO signature can be represented with the DR model of a magnetar by considering its precession motion and the Doppler boosting effect of its DR ejecta, assuming that the magnetar spindown is dominated by the GW emission and the DR ejecta is co-axial with the GRB jet axis. Adopting an efficiency of 0.05 for the DR wind, the bulk Lorentz factor of the DR wind is as constrained as $\gamma=8$, which is much lower than that of the GRB jet. The derived parameters of the magnetar are $B_{\rm p}= 7.9\times10^{13}$ Gauss, $P_{\rm s,0}= 0.98$ ms, $\epsilon = 6.3 \times 10^{-6}$, and $w=0.4$.

\begin{figure}
\center
\includegraphics[angle=0,scale=0.40]{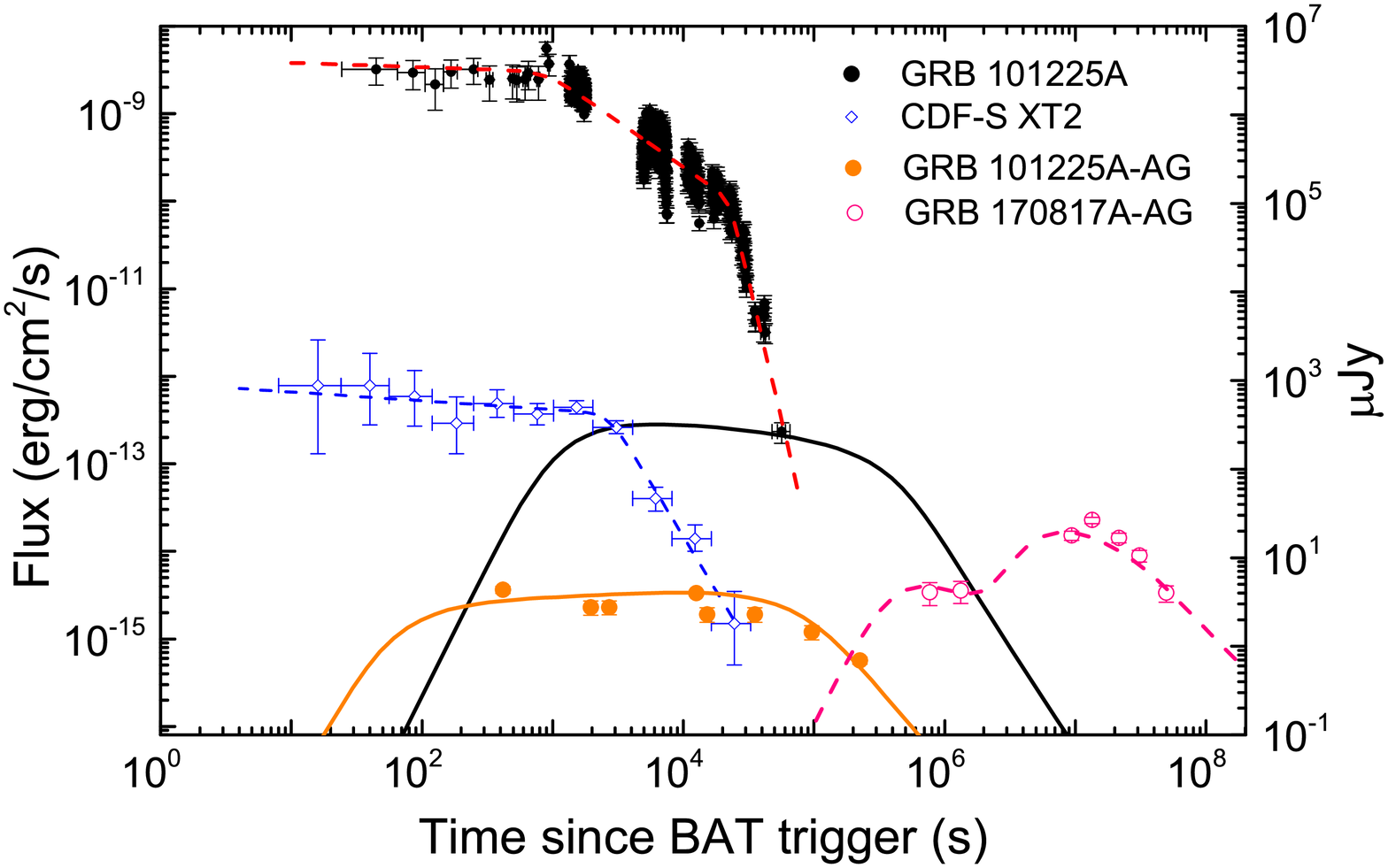}
\center\caption{Joint BAT+XRT (black dots) and R-band optical afterglow (orange dots) lightcurves of GRB 101225A. The global feature of the gamma-ray/X-ray lightcurve is depicted with a triple power-law function as marked with a red dashed line. The optical and X-ray emission of the external-forward shock derived from our model fit are shown with orange and black solid lines. The data of CDF-S XT2 is shown with blue open dots, with our empirical fit with a broken power-law function (blue dashed line). The X-ray afterglow data of GRB 170817A alongs with our fit and with an off-axis two-component jet model are also shown with pink open dots and pink dashed line.}
\label{1}
\end{figure}

\begin{figure}
\center
\includegraphics[angle=0,scale=0.30]{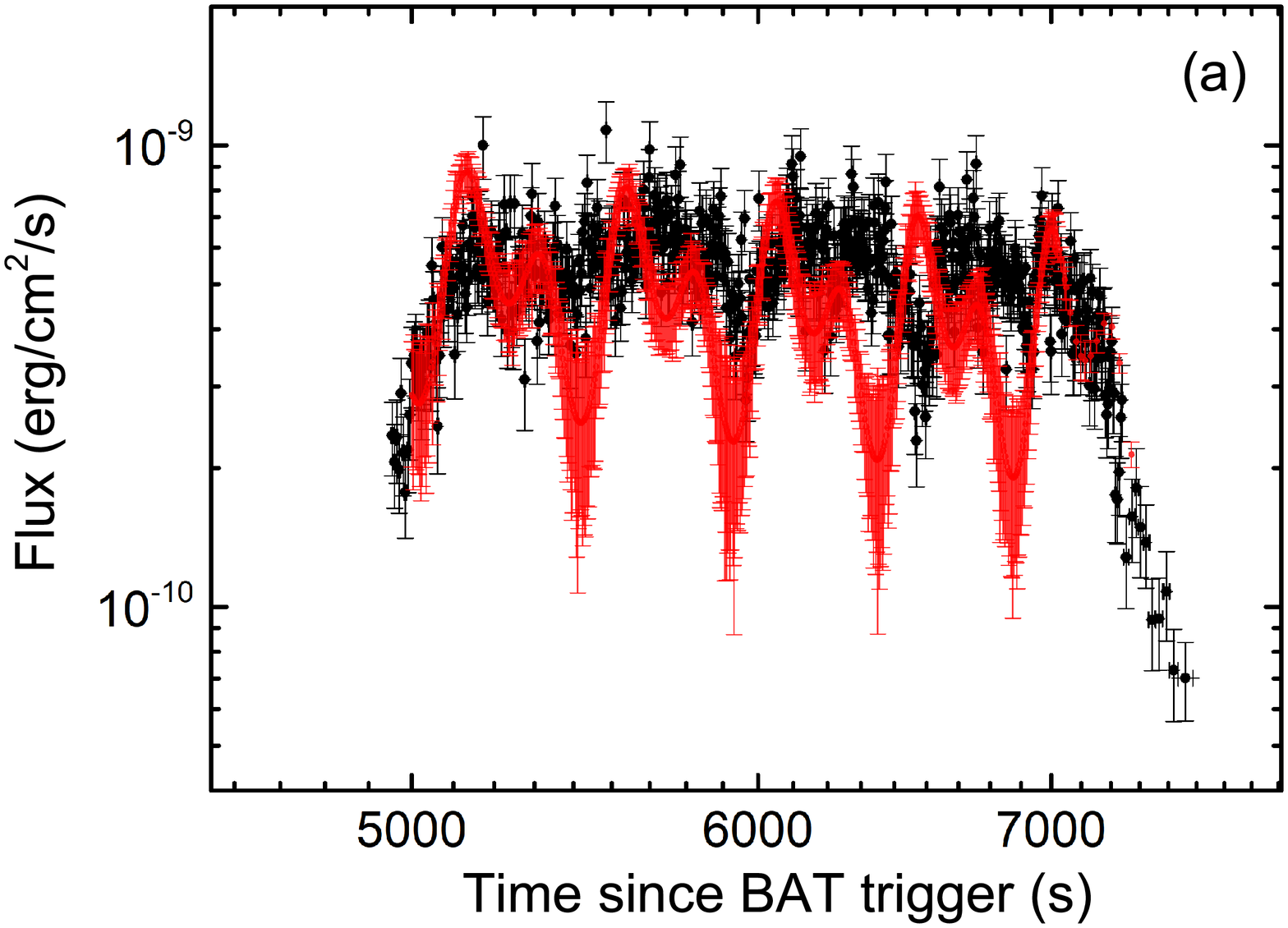}
\includegraphics[angle=0,scale=0.30]{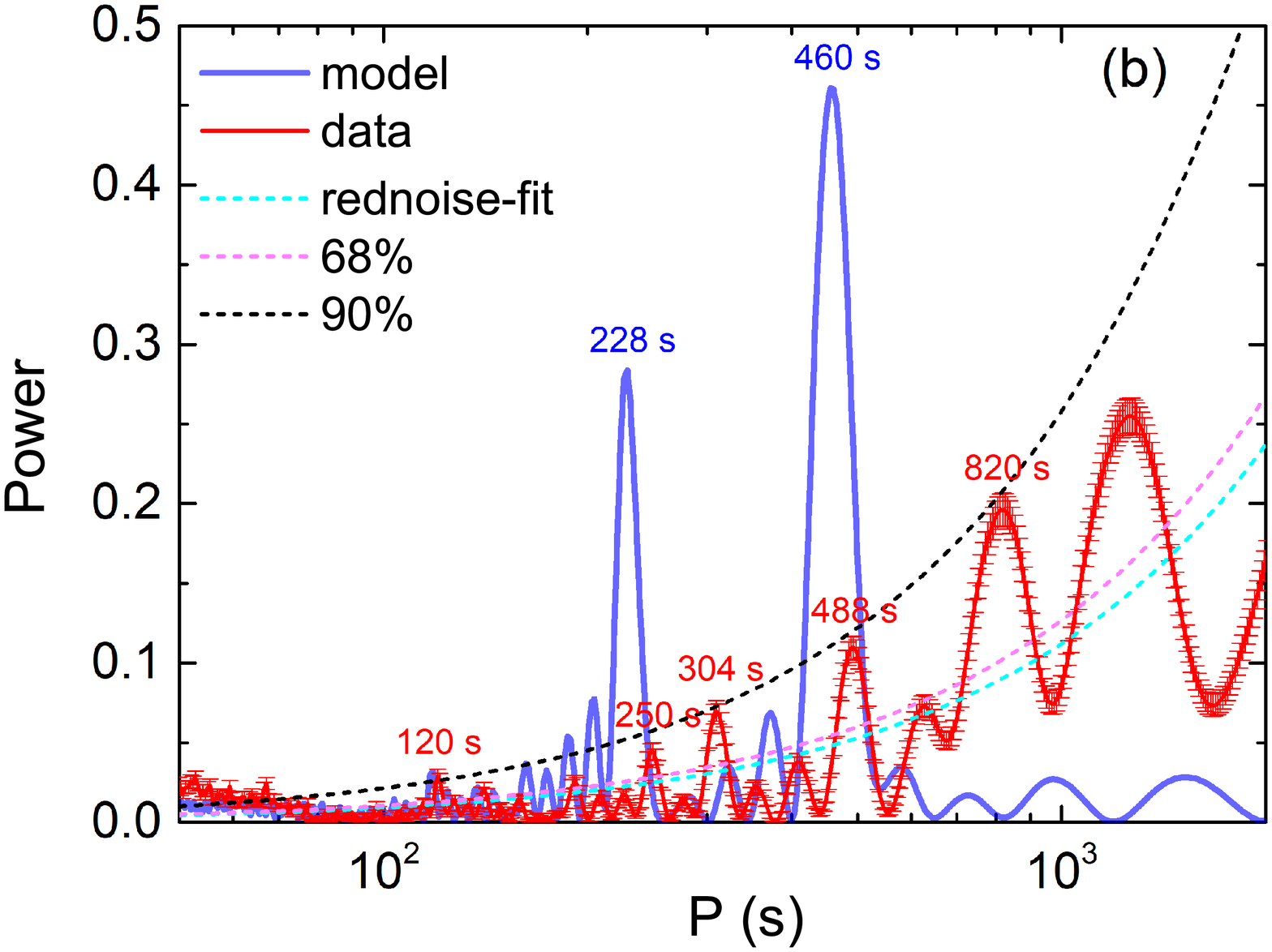}
\center\caption{{\em Left panel}---Observed X-ray lightcurve of GRB 101225A (black dots) in the time interval $[4900,\ 7500]$ s post the BAT trigger along with our theoretical lightcurve (red dots). {\em Right panel}--- The power density spectra (PDSs) calculated with the LSP algorithm for both the observed and theoretical lightcurves shown in the left panels. Our fit to the red-noise spectrum and its confidence levels of 68\% and 90\% are also shown.}
\label{2}
\end{figure}

\begin{figure}
\center
\includegraphics[angle=0,scale=0.4]{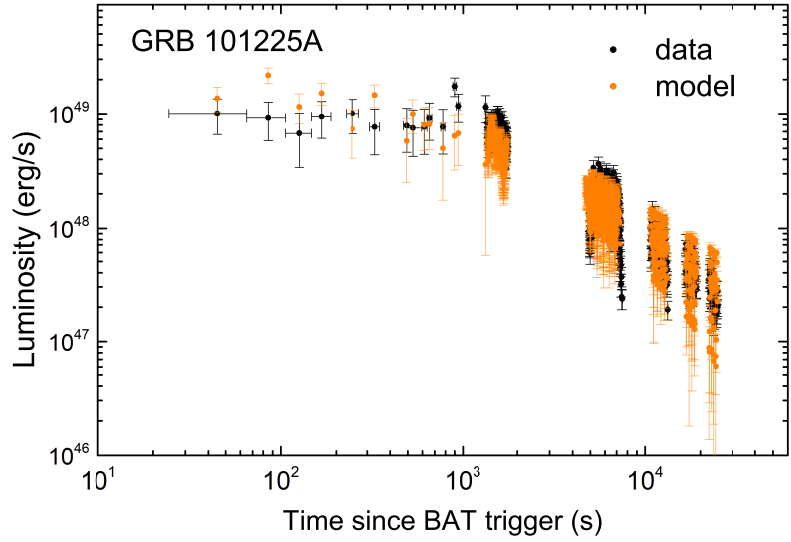}
\center\caption{The observed gamma-ray/X-ray lightcurve of GRB 101225A prior to $2.5\times 10^{4}$ (black dots) together with the theoretical lightcurve derived from our model (orange dots). The theoretical lightcurve is sampled the same as the data and the uncertainties of the theoretical points are also taken as the uncertainty of the corresponding data points. }
\label{3}
\end{figure}

GRB 170817A is believed to be an extremely off-axis GRB associated with GW 170817 (Takahashi \& Ioka 2021). We compare its X-ray afterglow lightcurve with GRB 101225A in Fig.\ref{1} and fit the lightcurve with a two-component jet model. Our fitting curve is also shown in Fig.\ref{1}. The model parameters are $\epsilon_e=8.3\times 10^{-3}$, $\epsilon_B=3.5\times 10^{-3}$, $p=2.1$, $n=3.9\times 10^{-5}$ cm$^{-3}$, $E_{\rm k, iso}=3.8\times 10^{52}$ erg, $\Gamma_{0}=120$, $\theta_{j}=2.9^{\rm o}$ for the narrow component and $\epsilon_e=5\times 10^{-2}$, $\epsilon_B=5\times 10^{-3}$, $p=2.24$, $n=3.9\times 10^{-5}$ cm$^{-3}$, $E_{\rm k, iso}=1.6\times 10^{50}$ erg, $\Gamma_{0}=6.4$, $\theta_{j}=28.7^{\rm o}$ for the wide component. Its geometrically-corrected jet energy is $7\times 10^{49}$ erg of GRB 170817A, which is comparable to that of GRB 101225A ($\sim 3\times 10^{50}$ erg). The derived viewing angle of GRB 170817A is $\theta_v=15.8^{\rm o}$, which is much larger than that of GRB 101225A ($3.7^{\rm o}$). The observed longer delay of the afterglow onset of GRB 170817A than GRB 101225A should be due to its larger viewing angle. Note that the initial Lorentz factors of the narrow and wide components are similar to the GRB jet and the DR ejecta of GRB 101225A derived in this analysis. It is possible that the nature of the two component jets is a combination of a co-axial GRB jet and DR ejecta, as shown in Xie et al. (2020).

GRB 101225A like X-ray transients (XT) is of great interest. Similar events were reported in literature. Using the data from the Deep Chandra Field-South (CDF-S) survey, Xue et al. (2019) reported an X-ray transient CDF-S XT2, which is associated with a galaxy at redshift $z = 0.738$. It lies on the outskirts of its star-forming host galaxy with a moderate offset from the galaxy center, favoring that the source is from the merger of two neutron stars (Xue et al. 2019; L\"{u} et al. 2021). Its X-ray flux keeps almost a constant, then decays with a temporal slope of -2.43 as shown in Fig.\ref{1} in comparison with GRB 101225A. The decay slope post the plateau is much steeper than that observed in GRB 101225A ($-1.1$), and even steeper than the prediction in the case that the EM emission dominates the spindown of the magnetar ($-2$). More recently, Lin et al. (2021) reported a similar event (XRT 210423) with the Chandra X-Ray Telescope. It starts with a fast rise within a few tens of seconds, then keeps a flux of $4\times 10^{-13}$ erg s$^{-1}$ cm$^{-2}$ to $4.1\times 10^3$ seconds, followed by a steep decay of $t^{-3.6}$. The X-ray lightcurves of CDF-S XT2 and XRT 210423 are analog to GRB 101225A, although they are much dimmer than GRB 101225A. GRB 101225A would be among the brightest ones of this kind of fast X-ray transients. It was proposed that such a population of X-ray transients may be observed with {\em Swift}/XRT (Xie et al. 2020). They would be promising EM counterparts of GW events and the associated kilonovaes if the magnetars are born in merger of two neutron stars.

In our analysis, we model the X-ray lightcurve of GRB 101225A in the scenario that the spin down of the magnetar is due to its rotation energy lost via the GW emission. This would set a lower limit of the energy release of the GW emission ($E_{\rm GW}$) as $E_{\rm GW}\geq E_{\rm EM}$, where $E_{\rm EM}$ is the kinetic energy of the DR wind. $E_{\rm GW}$ can be calculated by integrating $L_{\rm EM}$ over the time interval before the collapse of the magnetar. Based on our above analysis, we have $E_{\rm EM}\sim 4\times 10^{50}$ erg. Adopting $\epsilon=6.3\times 10^{-6}$ and $\Omega_0=6408$ $\rm rad$ $\rm s^{-1}$, we set the lower limit of $I$ with the condition of $E_{\rm GW}\geq E_{\rm EM}$. We get $I>6.35\times 10^{46}$ g $\rm cm^{2}$. The total rotational energy of the magnetar then is $E_{\rm rot}>2.0 \times 10^{53}$ erg.

We estimate the detection probability of its GW emission with LIGO. The strain of GW emission at frequency $f$ for a rotating neutron star at distance $D_{L}$ is given by

\begin{eqnarray}\label{h(t)}
h(t) = \frac{4GI\epsilon}{D_{L}c^{4}}\Omega(t)^{2}, \Omega(t)=2\pi f(t),
\end{eqnarray}

The optimal matched filter signal-to-noise ratio can be expressed as (Corsi \& M\'{e}sz\'{a}ros 2009)

\begin{eqnarray}\label{S/N}
S/N = \int^{+\infty}_{0}\left({\frac{h_{c}}{
h_{\rm rms}}}\right)^{2}d(\ln f),
\end{eqnarray}
where $h_{c}=fh(t)\sqrt{|dt/df|}$ is the characteristic amplitude, $h_{\rm rms}=\sqrt{fS_{h}(f)}$, and $dt=\tau_{\rm sd}d\Omega^{-4}/\Omega^{-4}_{0}$. In the limit, the characteristic amplitude of GW from a rotating NS can be estimated as (e.g. Dall'Osso \& Stella 2007; Corsi \& M\'{e}sz\'{a}ros 2009; Dall'Osso et al. 2009; Dall'Osso et al. 2015; Lasky \& Glampedakis 2016; Dall'Osso et al. 2018)

\begin{eqnarray}\label{h_{c}}
h_{c} (f) &= & fh(t)\frac{2\Omega^{2}_{0}}{\pi^{2}}(\tau_{\rm sd})^{\frac{1}{2}}f^{-\frac{5}{2}}\\
&\approx & 6.42\times 10^{-28}\left(\frac{\varepsilon}{10^{-5}}\right) \left(\frac{I}{10^{45}{\rm g\ cm}^{2}}\right)\left(\frac{D_{L}}{100{\rm Mpc}}\right)^{-1}\left(\frac{\Omega_{0}}{1000}\right)^{2}\left(\frac{\tau_{\rm sd}}{1000}\right)^{1/2}\left(\frac{f}{1{\rm kHz}}\right)^{1/2}.
\end{eqnarray}

\begin{figure}
\center
\includegraphics[angle=0,scale=0.40]{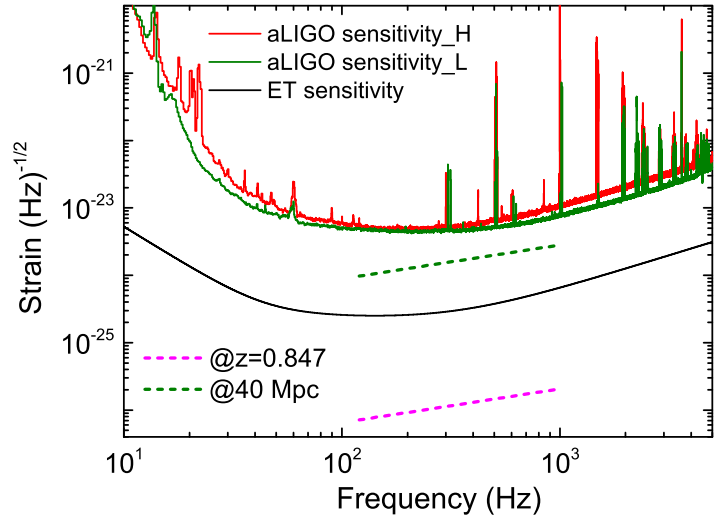}
\center\caption{Examination of the delectability of the GW emission from GRB 101225A with aLIGO and ET. The black solid line is the sensitivity limits for ET, and the red and dark cyan solid lines are the sensitivity limits for aLIGO-Hanford and aLIGO-Livingston, respectively.}
\label{4}
\end{figure}

We calculate $h_c(f)$ from $f = 120$ Hz to $1000$ Hz (aLIGO detector sensitive band) by adopting $I=6.35\times 10^{46}$ g cm$^{2}$. Our results are shown in Fig.\ref{4}. One can find that the GW emission cannot be detectable even if it is at a distance of 40 Mpc with current aLIGO and could be detected by the future ET.

One caveat should be addressed is that the moment of inertia of the magnetar and rational energy of the magnetar in GRB 101225A inferred from the conditions of $E_{\rm GW}\geq E_{\rm EM}$ and pure-GW dominated spindown violates the maximum moment of inertia and spin energy consistent with any NS equation of state (EoS). The inferred $E_{\rm rot}$ is $\sim2\times 10^{53}$ erg, which is even larger than the maximum NS spin energy given by the dynamical bar mode instability, i.e. $E_{\rm rot} \lesssim 1\times 10^{53}$ erg (Chandrasekhar 1969; Dall'Osso et al. 2018). In addition, by adopting $I=6.35\times 10^{46}$ g cm$^{2}$, the inferred spin-down timescale for the magnetar would be $3.33\times 10^{6}$ s, which is inconsistent with the observed one, i.e. $t_{\rm sd}=2.5\times 10^4$ seconds. Note that the parameters derived in this analysis are from a pure GW-driven spindown scenario. It is possible that the spindown is not purely GW-dominated and the moment of inertia may be overestimated.

\section{Acknowledgements}

We very appreciate helpful comments and suggestions from the referee. We thank Cong Yu, Hou-Jun L\"{u}, Qin-He Yang, and Kuan Liu for useful discussion. We acknowledge the use of the public data from the {\em Swift} data archive and the UK {\em Swift} Science Data Center. This work is supported by the National Natural Science Foundation of China (Grant No.12133003 and U1731239), Guangxi Science Foundation (grant No. 2017AD22006), and Innovation Project of Guangxi Graduate Education (YCBZ2020025).

{}

\clearpage

\begin{thebibliography}{}

\bibitem[Abbott et al.(2017)]{2017PhRvL.119p1101A} Abbott, B.~P., Abbott, R., Abbott, T.~D., et al.\ 2017, \prl, 119, 161101. doi:10.1103/PhysRevLett.119.161101
\bibitem[Abbott et al.(2017)]{2017ApJ...848L..13A} Abbott, B.~P., Abbott, R., Abbott, T.~D., et al.\ 2017, \apjl, 848, L13. doi:10.3847/2041-8213/aa920c
\bibitem[Arzamasskiy et al.(2015)]{2015MNRAS.453.3540A} Arzamasskiy, L., Philippov, A., \& Tchekhovskoy, A.\ 2015, \mnras, 453, 3540. doi:10.1093/mnras/stv1818
\bibitem[Campana et al.(2011)]{2011Natur.480...69C} Campana, S., Lodato, G., D'Avanzo, P., et al.\ 2011, \nat, 480, 69. doi:10.1038/nature10592
\bibitem[Chandrasekhar(1969)]{1969efe..book.....C} Chandrasekhar, S.\ 1969, The Silliman Foundation Lectures, New Haven: Yale University Press, 1969
\bibitem[Cummings \& Sakamoto(2010)]{2010GCN.11504....1C} Cummings, J.~R. \& Sakamoto, T.\ 2010, GRB Coordinates Network, Circular Service, No. 11504, \#1 (2010), 11504
\bibitem[Cutler(2002)]{2002PhRvD..66h4025C} Cutler, C.\ 2002, \prd, 66, 084025. doi:10.1103/PhysRevD.66.084025
\bibitem[Dai \& Lu(1998)]{1998PhRvL..81.4301D} Dai, Z.~G. \& Lu, T.\ 1998, \prl, 81, 4301. doi:10.1103/PhysRevLett.81.4301
\bibitem[Dall'Osso et al.(2009)]{2009MNRAS.398.1869D} Dall'Osso, S., Shore, S.~N., \& Stella, L.\ 2009, \mnras, 398, 1869. doi:10.1111/j.1365-2966.2008.14054.x
\bibitem[Dall'Osso \& Stella(2007)]{2007Ap&SS.308..119D} Dall'Osso, S. \& Stella, L.\ 2007, \apss, 308, 119. doi:10.1007/s10509-007-9323-0
\bibitem[Dall'Osso et al.(2018)]{2018MNRAS.480.1353D} Dall'Osso, S., Stella, L., \& Palomba, C.\ 2018, \mnras, 480, 1353. doi:10.1093/mnras/sty1706
\bibitem[Dall'Osso \& Stella(2021)]{2021arXiv210310878D} Dall'Osso, S. \& Stella, L.\ 2021, arXiv:2103.10878
\bibitem[Dall'Osso et al.(2015)]{2015ApJ...798...25D} Dall'Osso, S., Giacomazzo, B., Perna, R., et al.\ 2015, \apj, 798, 25. doi:10.1088/0004-637X/798/1/25
\bibitem[Falcke \& Rezzolla(2014)]{2014A&A...562A.137F} Falcke, H. \& Rezzolla, L.\ 2014, \aap, 562, A137. doi:10.1051/0004-6361/201321996
\bibitem[Goldreich(1970)]{1970ApJ...160L..11G} Goldreich, P.\ 1970, \apjl, 160, L11. doi:10.1086/180513
\bibitem[Goldreich \& Julian(1969)]{1969ApJ...157..869G} Goldreich, P. \& Julian, W.~H.\ 1969, \apj, 157, 869. doi:10.1086/150119
\bibitem[Haskell et al.(2008)]{2008MNRAS.385..531H} Haskell, B., Samuelsson, L., Glampedakis, K., et al.\ 2008, \mnras, 385, 531. doi:10.1111/j.1365-2966.2008.12861.x
\bibitem[Jaranowski et al.(1998)]{1998PhRvD..58f3001J} Jaranowski, P., Kr{\'o}lak, A., \& Schutz, B.~F.\ 1998, \prd, 58, 063001. doi:10.1103/PhysRevD.58.063001
\bibitem[Jones(1976)]{1976Ap&SS..45..369J} Jones, P.~B.\ 1976, \apss, 45, 369. doi:10.1007/BF00642671
\bibitem[Kalapotharakos \& Contopoulos(2009)]{2009A&A...496..495K} Kalapotharakos, C. \& Contopoulos, I.\ 2009, \aap, 496, 495. doi:10.1051/0004-6361:200810281
\bibitem[Levan et al.(2014)]{2014ApJ...781...13L} Levan, A.~J., Tanvir, N.~R., Starling, R.~L.~C., et al.\ 2014, \apj, 781, 13. doi:10.1088/0004-637X/781/1/13
\bibitem[Liang et al.(2007)]{2007ApJ...670..565L} Liang, E.-W., Zhang, B.-B., \& Zhang, B.\ 2007, \apj, 670, 565. doi:10.1086/521870
\bibitem[Lin et al.(2021)]{2021ATEL...14599} Lin, D.-C, Jimmy A.~I., Edo B.\ 2021,
    ATEL, 14599.
\bibitem[Lomb(1976)]{1976Ap&SS..39..447L} Lomb, N.~R.\ 1976, \apss, 39, 447. doi:10.1007/BF00648343
\bibitem[L{\"u} et al.(2018)]{2018MNRAS.480.4402L} L{\"u}, H.-J., Zou, L., Lan, L., et al.\ 2018, \mnras, 480, 4402. doi:10.1093/mnras/sty2176
\bibitem[L{\"u} \& Zhang(2014)]{2014ApJ...785...74L} L{\"u}, H.-J. \& Zhang, B.\ 2014, \apj, 785, 74. doi:10.1088/0004-637X/785/1/74
\bibitem[Margutti et al.(2017)]{2017ApJ...848L..20M} Margutti, R., Berger, E., Fong, W., et al.\ 2017, \apjl, 848, L20. doi:10.3847/2041-8213/aa9057
\bibitem[Metzger et al.(2011)]{2011MNRAS.413.2031M} Metzger, B.~D., Giannios, D., Thompson, T.~A., et al.\ 2011, \mnras, 413, 2031. doi:10.1111/j.1365-2966.2011.18280.x
\bibitem[Metzger et al.(2018)]{2018ApJ...856..101M} Metzger, B.~D., Thompson, T.~A., \& Quataert, E.\ 2018, \apj, 856, 101. doi:10.3847/1538-4357/aab095
\bibitem[Philippov et al.(2014)]{2014MNRAS.441.1879P} Philippov, A., Tchekhovskoy, A., \& Li, J.~G.\ 2014, \mnras, 441, 1879. doi:10.1093/mnras/stu591
\bibitem[Philippov et al.(2015)]{2015ApJ...801L..19P} Philippov, A.~A., Spitkovsky, A., \& Cerutti, B.\ 2015, \apjl, 801, L19. doi:10.1088/2041-8205/801/1/L19
\bibitem[Ren et al.(2020)]{2020ApJ...901L..26R} Ren, J., Lin, D.-B., Zhang, L.-L., et al.\ 2020, \apjl, 901, L26. doi:10.3847/2041-8213/abb672
\bibitem[Scargle(1982)]{1982ApJ...263..835S} Scargle, J.~D.\ 1982, \apj, 263, 835. doi:10.1086/160554
\bibitem[Spitkovsky(2006)]{2006ApJ...648L..51S} Spitkovsky, A.\ 2006, \apjl, 648, L51. doi:10.1086/507518
\bibitem[Stella et al.(2005)]{2005ApJ...634L.165S} Stella, L., Dall'Osso, S., Israel, G.~L., et al.\ 2005, \apjl, 634, L165. doi:10.1086/498685
\bibitem[Suvorov \& Kokkotas(2020)]{2020ApJ...892L..34S} Suvorov, A.~G. \& Kokkotas, K.~D.\ 2020, \apjl, 892, L34. doi:10.3847/2041-8213/ab8296
\bibitem[Takahashi \& Ioka(2021)]{2021MNRAS.501.5746T} Takahashi, K. \& Ioka, K.\ 2021, \mnras, 501, 5746. doi:10.1093/mnras/stab032
\bibitem[Th{\"o}ne et al.(2011)]{2011Natur.480...72T} Th{\"o}ne, C.~C., de Ugarte Postigo, A., Fryer, C.~L., et al.\ 2011, \nat, 480, 72. doi:10.1038/nature10611
\bibitem[Troja et al.(2007)]{2007ApJ...665..599T} Troja, E., Cusumano, G., O'Brien, P.~T., et al.\ 2007, \apj, 665, 599. doi:10.1086/519450
\bibitem[Usov(1992)]{1992Natur.357..472U} Usov, V.~V.\ 1992, \nat, 357, 472. doi:10.1038/357472a0
\bibitem[Vaughan(2005)]{2005A&A...431..391V} Vaughan, S.\ 2005, \aap, 431, 391. doi:10.1051/0004-6361:20041453
\bibitem[Vaughan(2010)]{2010MNRAS.402..307V} Vaughan, S.\ 2010, \mnras, 402, 307. doi:10.1111/j.1365-2966.2009.15868.x
\bibitem[Xiao \& Dai(2019)]{2019ApJ...878...62X} Xiao, D. \& Dai, Z.-G.\ 2019, \apj, 878, 62. doi:10.3847/1538-4357/ab12da
\bibitem[Xie et al.(2020)]{2020ApJ...894...52X} Xie, W.-J., Zou, L., Liu, H.-B., et al.\ 2020, \apj, 894, 52. doi:10.3847/1538-4357/ab8302
\bibitem[Xue et al.(2019)]{2019Natur.568..198X} Xue, Y.~Q., Zheng, X.~C., Li, Y., et al.\ 2019, \nat, 568, 198. doi:10.1038/s41586-019-1079-5
\bibitem[Zanazzi \& Lai(2015)]{2015MNRAS.451..695Z} Zanazzi, J.~J. \& Lai, D.\ 2015, \mnras, 451, 695. doi:10.1093/mnras/stv955
\bibitem[Zhang(2014)]{2014ApJ...780L..21Z} Zhang, B.\ 2014, \apjl, 780, L21. doi:10.1088/2041-8205/780/2/L21
\bibitem[Zhang \& M{\'e}sz{\'a}ros(2001)]{2001ApJ...552L..35Z} Zhang, B. \& M{\'e}sz{\'a}ros, P.\ 2001, \apjl, 552, L35. doi:10.1086/320255
\bibitem[Zou et al.(2019)]{2019ApJ...877..153Z} Zou, L., Zhou, Z.-M., Xie, L., et al.\ 2019, \apj, 877, 153. doi:10.3847/1538-4357/ab17dc
\bibitem[{\c{S}}a{\c{s}}maz Mu{\c{s}} et al.(2019)]{2019ApJ...886....5S} {\c{S}}a{\c{s}}maz Mu{\c{s}}, S., {\c{C}}{\i}k{\i}nto{\u{g}}lu, S., Ayg{\"u}n, U., et al.\ 2019, \apj, 886, 5. doi:10.3847/1538-4357/ab498c


\end{thebibliography}
\end{document}